**Graphene: Piecing it together**

By *Mark H. Rümmeli\*, Claudia G. Rocha, Frank Ortmann, Imad Ibrahim, Haldun Sevincli, Felix Börrnert, Jens Kunstmann, Alicja Bachmatiuk, Markus Pötsche, Masashi Shiraishi, Meyya Meyyappan, Bernd Büchner, Stephan Roche and Gianaurelio Cuniberti\**

[*]     Prof. Dr. G. Cuniberti
Institute for Materials Science and Max Bergmann Center of Biomaterials, Dresden University of Technology, Dresden 01062, Germany and Division of IT Convergence Engineering, POSTECH, Pohang 790-784, Republic of Korea
E-mail: g.cuniberti@tu-dresden.de
        Dr. M. H. Rümmeli
Leibniz-Institut für Festkörper- und Werkstoffforschung Dresden e. V., PF 27 01 16, 01171 Dresden, Germany and Technische Universität Dresden, 01062 Dresden, Germany
Email:m.ruemmeli@ifw-dresden.de
        Dr. C. G. Rocha, Dr. H. Sevincli, Dr. J. Kunstmann, M. Pötsche
Institute for Materials Science and Max Bergmann Center of Biomaterials, Dresden University of Technology, Dresden 01062, Germany
        Dr. F. Ortmann
INAC/SPrAM, CEA Grenoble, 17 Rue des Martyrs, 38054 Grenoble, France
        I. Ibrahim
Leibniz-Institut für Festkörper- und Werkstoffforschung Dresden e. V., PF 27 01 16, 01171 Dresden, Germany and Technische Universität Dresden, 01062 Dresden, Germany
        F. Börrnert, Dr. A Bachmatiuk, Prof. Dr. B. Büchner
Leibniz-Institut für Festkörper- und Werkstoffforschung Dresden e. V., PF 27 01 16, 01171 Dresden, Germany
        Prof. Dr. M. Shiraishi
Graduate School of Engineering Science, Osaka University, Machikaneyama-cho 1-3, Toyonaka-shi 560-8531, Osaka, Japan
        Dr. M. Meyyappan
Center for Nanotechnology, NASA Ames Research Center, Moffett Field, US; Division of IT Convergence Engineering, POSTECH, Pohang 790-784, Republic of Korea
        Prof. Dr. S. Roche
Catalan Institute of Nanotechnology, Barcelona, Spain.

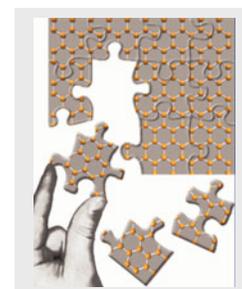

Graphene has a multitude of striking properties that make it an exceedingly attractive material for various applications,   many of which will emerge over the next decade. However, one of the most promising applications lie in exploiting its peculiar electronic properties which are governed by its electrons obeying a linear dispersion relation. This leads to the observation of half integer quantum hall effect and the absence of localization. The latter is attractive for graphene-based field effect transistors. However, if graphene is to be the material for future electronics, then significant hurdles need to be surmounted, namely, it needs to be mass produced in an economically viable manner and be of high crystalline quality with no or virtually no defects or grains boundaries. Moreover, it will need to be processable with atomic precision. Hence, the future of graphene as a material for electronic based devices will depend heavily on our ability to piece graphene together as a single crystal and define its edges with atomic precision. In this progress report, the properties of graphene that make it so attractive as a material for electronics is introduced to the reader. The focus then centers on current synthesis strategies for graphene and their weaknesses in terms of electronics applications are highlighted.

1. Introduction

Early theoretical works [1] suggested that pristine two dimensional (2D) crystals like graphene would be unstable due to thermal fluctuations preventing long range crystalline order at finite temperatures. The ground breaking works of Geim and Novoselov [2] not only showed graphene could exist as a single layer of carbon atoms structured in a honeycomb lattice, but that it has intriguing properties. Since then an abundance of new physics has been and continues to be revealed. This is reflected in the exponential increase in publications on graphene as illustrated in figure 1.

In terms of fundamental physics, graphene has already proven itself. The plethora of physical properties has opened numerous potential applications in disparate fields, some of which are likely to succeed within a decade. These include graphene based composites, supercapacitors, batteries, fuel cells, conductive pads and inks, touch screen displays which can also be flexible, generic intelligent coatings and sensors. One of the most relevant and exciting application potential of graphene is as a base material for future electronic devices (high frequency and analogue). However, in order to realize this potential significant hurdles in the preparation and processing of graphene need to be surmounted. This is because graphene devices, on the whole, require reliable and reproducible fabrication of defect-free graphene with atomic precision in large volume. Current synthesis strategies are not sufficiently developed for high volume graphene fabrication, let alone with atomic precision. A significant concern is the presence of grains and grain boundaries in synthesized graphene (see figure 2).[3] While some routes can minimize this problem, for example through growth over single crystals, these methods are not suited to mass production. The future of graphene as a material for electronic and spin based devices will depend heavily on our ability to piece graphene together as a single crystal and define its edges with atomic precision. The first part of this progress report briefly introduces the reader to the fundamental properties of graphene that make it so attractive for electronic device fabrication. The focus then turns to current synthesis strategies for graphene and some of its derivatives. Their weaknesses in terms of suitability for device fabrication are highlighted.

2. The peculiar electronic properties and transport characteristics of graphene

The unusual properties of graphene such as its low-energy excitation spectrum given by a linear energy dispersion [4] stand in stark contrast to the quadratic energy-momentum relation obeyed by electrons in conventional semiconductors (see figure 3). The electrons in graphene mimic the behavior of massless relativistic particles described by the Dirac equation with an effective Fermi velocity that is around 300 times less than the speed of the light. [2.b, 5] This makes graphene a promising system to study quantum electrodynamics phenomena, an area of investigation previously limited to particle physics and cosmology. Early experimental studies confirmed that suspended graphene had outstanding carrier mobilities approaching 200000 cm$^2$/Vs [6] as previously predicted [7] and mean free paths of the order of microns at room temperature. [6] Moreover, the charge carrier density of suspended graphene vanishes at the charge neutrality point; a minimum conductivity approaching the ballistic value on the order of $e^2/h$ (e being the electron charge and h Planck's constant) is found at low temperatures. [6, 8] When graphene is supported on substrates altered spatial fluctuations of the electrostatic potential lead to local Fermi level shifts about the Dirac point resulting in electron-hole puddles (regions of finite carrier density). [9] In fact, transport measurements average over many such puddles, resulting in energy-convoluted signals which limit direct access to the Dirac point.

The intrinsic electronic properties of pristine graphene are well understood and on their own would not cause much excitement. However, these properties interfere with intrinsic and extrinsic scatterers and give rise to novel transport phenomena which is a very active research field. The term disorder is used to distinguish sources of unintentional modifications on graphene's electronic and transport properties from intentional ones such as the introduction of an energy gap through substrate interaction. [10] Scattering centres can form via local fields induced by an underlying substrate [11] or by adsorbates, *viz.* atoms [12] and molecules [13] , or by modification of the 2D honeycomb structure. Zero dimensional defects are essentially topological defects, for example, the formation of pentagon/heptagons such as Stone-Wales defects, which for the most part retain $sp^2$ bonding character. Often heptagon-pentagon pairs accompany larger 1D defects such as small angle domain boundaries. [14] Other examples which also maintain the $sp^2$ hybridization character of graphene include vacancies, and dopants such as boron or nitrogen. Hydrogen or oxygen as a dopant leads to $sp^3$-type hybridization, which, when existing in high concentration forms a graphene derivative, namely, graphane and graphene oxide respectively.

Weak disorder, which still allows long mean free paths in the micron range, may give rise to novel transport phenomena such as weak localization (WL) (characterized by enhanced backscattering) or weak anti-localization (WAL) (see figure 4). [15] The latter has only been previously observed in systems with strong spin-orbit coupling. The light carbon atoms comprising graphene make this coupling negligible. In general, the electrical conductivity in low-dimensional weakly disordered metals may be described as a sum of a semiclassical (Drude) term and a quantum-correction due to multiple interfering partial waves. The sign of the quantum correction distinguishes weak localization (negative) from weak antilocalization behaviour (positive). Switching between WL and WAL has only been observed in ordinary metals and can be tuned with the strength of the spin-orbit coupling. [15.a] In graphene such tuning is possible without spin-orbit coupling as was first predicted theoretically [15.b, 16] and later demonstrated experimentally by Tikhonenko et al. [15.d] Different magneto-conductance responses are obtained depending on temperature (by tuning decoherence), magnetic field and Fermi level position. The suppression of backscattering in graphene in the presence of long-range disorder (such as the screening of charges by an underlying oxide) has been probed by magnetoconductance measurements under weak magnetic fields. [15.b,d, 17] Another intriguing feature in graphene, which remains controversial, concerns the dc-conductivity minimum at the Dirac point. The concept of a minimum conductivity assumes that the above discussed weak localization effects can be neglected. A number of early theoretical works agree on a value of $\sigma=4e^2/h$ [18] for short range scatters that are relatively insensitive to concentration in the weak disorder regime. [8.c] Experimentally larger values are obtained ($4e^2/h$ or higher). [8.a, 19] Despite confirmation by other groups, [20] due to deviations between samples a universal experimental minimum has not been demonstrated thus far.

This discrepancy between theory and experiment may be related to the existence of electron-hole puddles. [9, 11] Electron hole puddles raise questions on the energy dependence of the conductivity. In experiment, the energy dependent conductivity is often linear for small energies [2.b, 21] or slightly sublinear. [20] Different theoretical models are able to explain an approximate ~n behaviour for strong scatters [18.d] or ~n behaviour assuming long-range scattering potentials from charged impurities. [11, 18.e] On the other hand, short-range (wavevector-independent) potentials, which mimic dopants or vacancies, are expected to have a logarithmic or similar sublinear behaviour in the weak scattering limit. [18.a,b,d,e] Thus, significant efforts to better understand the impact of intentional doping, for example, with boron, nitrogen, oxygen, fluorine, or hydrogen are being conducted. [22] Theoretical studies reveal an electron/hole asymmetry for the conductivity, however, the conductance properties do not alter dramatically even with significant doping concentrations. [22.b] In addition to single dopants, more complex doping systems are beginning to emerge. A recent example is the incorporation of hexagonal BN domains within graphene. [23] The resultant BN-doped graphene is resistive. Graphene oxide [24] is derived from oxygen-functionalized graphene which occurs naturally and can be easily synthesized. Its properties, with sufficient oxygen doping (40-80%), [25] are very different from graphene. This is seen for a variety of mechanical and optoelectronic properties showing a large variance induced by different reduction levels. The conductance properties range from insulating to semi-metallic while similarly high optical transmittance can be observed. [26] Thus, it is argued that patterned graphene oxide on graphene may be a means to implement circuits. Obviously this will require a degree of control in patterning the oxide regions.

2.1 Tailoring graphene's boundaries: nanoribbons

Besides the idealizations of graphene-like 2D membranes, atomistic models of thin graphene strips – known as graphene nanoribbons (GNR) are also of great interest, in particular the nature of edge dislocations and the appearance of defective dangling bonds in carbon networks. [27] The ease with which graphene can be synthesized in a free state, combined with modern lithography techniques, has demonstrated that graphene can be structured sufficiently small so as to induce confinement effects (e.g. band gap opening). The physical properties of graphene nanoribbons are highly dependent on their width and the topology of the edge structures. [28] There are two canonical types of graphene ribbons referred to as armchair (AGNR) and zigzag (ZGNR) ribbons in accordance with their atomic arrangement along the edges (see figure 5).

According to the single-π band tight binding description with nearest-neighbor interactions, one can predict that zigzag ribbons, regardless of width, show a singular edge state which decays exponentially into the center of the ribbon. These edge states are twofold degenerate at the Fermi energy and reveal a non-dispersive feature which lasts about 1/3 of the total size of the graphene Brillouin Zone. As a consequence, the density of states of zigzag ribbons is characterized by a pronounced peak located at the charge neutrality point. In stark contrast, no such localized state appears in nanoribbons having an armchair edge configuration. Armchair ribbons have a rather sensitive electronic structure with respect to their width (see figure 6). [29] The dominant scattering processes and resulting transport features of graphene structures are very dependent on the range of the disorder potential and the robustness or destruction of the underlying sublattice symmetries. In particular, graphene nanoribbons are naturally subjected to edge disorder due to the high reactivity of edges that can be subjected to chemical passivation, roughness and structural reconstruction. Depending on the edge-shape of the graphene structure, different band gaps for similarly sized systems can be generated since its electronic structure is greatly influenced by disorder. [30] Transport measurements realized with etched graphene samples have demonstrated that sharp resonances can appear inside the transport gap, evidencing that the atomic details of tailored graphene systems play an important role on their conducting properties. [31] Atomistic models for edge disorder indicate that even when very weak edge irregularities are simulated, a prominent modification in the conductance profile of the nanoribbons is obtained. [32] Similar studies performed in disordered armchair ribbons have shown that they are relatively more sensitive to edge-disorder in comparison to zigzag configurations. [33] The effect of edge-disorder shape in ribbons also plays an important role on their thermal conductivity. The propagation of heat can be strongly suppressed as the irregularities along their edges become more pronounced. [34]

Several other perturbing agents can be used to control the electronic transmission of graphene nanoribbons, thus opening a wide field of technological implementations for these materials. External magnetic/electrical fields, [35] doping, [36] and multi-layer stacking [37] are some of the strategies adopted to manipulate the electronic response of graphene ribbons. Recent studies targeting the use of time-dependent fields in carbon-based materials [38] shed light on this growing research area, often overshadowed by studies considering external fields in the stationary regime. Under ac signals, several theoretical works have highlighted graphene's potential as a spectrometer device that can even operate with high frequency noise. In particular, for the case in which a homogeneous ac gate can act on graphene channels, it has been shown that it is possible to achieve full control of the conductance patterns which, remarkably, resemble Fabry-Perot interference patterns of light wave cavities. [39]

The outstanding mechanical properties of graphene have also attracted the interest from electronic applications due to the potential use that these light, stiff and flexible materials can offer for designing building block components in nanoelectromechanical systems (NEMS). According to several theoretical works, the electronic structure of suspended graphene is extremely resistant against mechanical forces, being able to support reversible elastic deformations above 20%. [40] Band gap engineering in response to mechanical strain in graphene is possible when the membranes are reduced to nanoribbons. [41]

In addition to such amazing electronic properties, graphene nanoribbons, in particular those with zigzag edges, are believed to give rise to magnetic instability associated to the pronounced flat band located at the Fermi energy ($E_F$). [27, 42] According to numerous theoretical studies based on density functional theory or on the mean-field Hubbard model, the pronounced peak in the density of states at the Dirac point found for zigzag ribbons can be split into an energy gap when spin-polarized states are considered. The magnetic configuration of system's ground state is a ferromagnetic alignment along the axial direction and antiferromagnetic orientation between opposite edges. In this sense, numerous suggestions indicating the use of magnetic-induced edges in graphene strips for spintronics applications came to light. [43] However, magnetic edges have never been observed experimentally. There is only an indirect observation reported recently by V.L.J. Joly et al. performed via local probe microscopy. [44] The difficulty in observing magnetic edge states is supported by recent theoretical works indicating that the ideal zigzag edge morphology is not very likely to exist. In fact other nonmagnetic edge morphologies are thermodynamically much more favourable. [45] Moreover, it was recently shown that even in perfect ZGNR the antiferromagnetic ground state is only stable at very low temperatures, [45.c] putting in check whether the phenomenon of edge magnetism in graphene is applicable at room temperature or not.

2.2 Technology demonstrations

The interesting properties mentioned above have led to the exploration of numerous applications such as transistors, chemical and biosensors, energy storage devices, nanoelectromechanical systems and others, just as the research community has done with carbon nanotubes previously.   There have been numerous investigations on the potential use of graphene as the conducting channel in CMOS transistors and also in novel Beyond-CMOS type architectures. [46]   As mentioned earlier, opening up a bandgap is the first big challenge to realize useful devices.  Various avenues pursued to achieve this goal have included introduction of uniaxial strain on the graphene layer, exploiting graphene-substrate interaction, creation of nanoribbons, and breaking the inversion symmetry in bilayer graphene by applying vertical electric fields.  Once this critical problem is solved, large scale device fabrication will need to address issues related to controlled growth on substrates, choice of substrate, contact engineering,  doping, high-κ gate dielectric and others.

Recently, Xia et al [46.c] obtained a room temperature on-off ratio of 100 for a bilayer graphene field effect transistor (GFET). The same group also demonstrated [46.d] a cut-off frequency as high as 100 GHz with one to two layers of graphene epitaxially formed on a 5 cm wafer. The devices exhibited a carrier density of 3 x $10^{12}$ $cm^{-2}$ and a Hall mobility of 1000-1500 $cm^2$/V.s. While most of the works transferred graphene layers grown elsewhere in GFET fabrication, Kondo et al. demonstrated an in situ, patterned growth of graphene by CVD at 650° C using a 200 nm thick Fe catalyst. [46.e] This catalyst layer was later removed by wet etching after the formation of the source and drain contacts. A current density of 108 A/$cm^2$ and device transconductance of 22 mS/mm were obtained for a channel length of 3 μm. Kim et al [46.f] fabricated a dual-gate GFET with $Al_2O_3$ as gate dielectric, however the graphene layer grown on copper had to be transferred to the device platform as in many other efforts in the literature. Their devices yielded a mobility of 8000 $cm^2$/V.s. Liao et al [46.g] also used $Al_2O_3$ and showed that the contrast of single layer graphene on 72 nm $Al_2O_3$/Si substrate is much better than that on 300 nm $SiO_2$/Si substrate. The higher dielectric constant of $Al_2O_3$ with smaller thickness of the dielectric also allowed a transconductance 7 times higher than the results for $SiO_2$.

Sub-10 nm wide graphene nanoribbons were used in a bottom-gated GFET configuration which produced on-off ratios of 106 and an on-state current density as high as 200 µA/mm. [46.j] Moon et al [46.k] prepared epitaxial graphene on the Si-face of 6H-SiC substrated and fabricated rf-field effect transistors; their devices showed an extrinsic transconductance of 148 MS/mm. Besides logic applications, graphene field effect devices have been considered for nonvolatile memory as well. [46.l] Good cyclability and a reset of 80 µs have been obtained with 2 µm channel width and 4 µm gate length devices.

Chemical sensors work primarily based on the change in resistance, capacitance or some other easily measurable property when a gas or vapor gets adsorbed on a conducting material. Graphene has been a candidate for chemical sensing in a number of investigations. [47]

Schedin et al. were able to detect changes in local carrier concentration one by one electron and thus a stepwise resistance change, upon adsorption of a vapor on a graphene layer. [47.a] Fowler et al. coated a single layer graphene film on a interdigitated electrode and showed conductance change upon adsorption of gas molecules such as $NO_2$, $NH_3$ and dinitrotoulene. [47.b] They postulated a charge transfer mechanism for the observed change in conductance of the graphene film. The detection limits in this work were 5 ppm, 5 ppm and 53 ppb respectively for the three gases above. Arsat et al fabricated a surface acoustic wave (SAW) based sensor using graphene as the conducting layer and assessed its sensing performance towards $H_2$ and CO. [47.c] The sensing response was measured to be 5.8 and 8.5 KHz for $H_2$ and 1000 ppm CO respectively. Lu et al also concluded that electron transfer between graphene and adsorbed gases such as $NH_3$ and $NO_2$ are responsible for their demonstration of detecting 1% and 100 ppm of these gases respectively. [47.d] Dua et al printed graphene onto flexible sheets by an ink-jet technique and used this chemiresistor to sense $NO_2$ and chlorine. [47.e]

Carbon based electrodes have been in continuous use for over hundred years including carbon black, glassy carbon, highly oriented pyrolytic graphite, carbon fiber, and more recently carbon nanotube electrodes. Nanoscale electrodes offer advantages in terms of reduced background noise because the biomolecules and electrodes are of comparable size, and the ability to work with small amount of target molecules relative to micro and macro electrodes. Graphene based electrodes have been investigated for biosensing applications. [48] Alwarappan et al [48.a] demonstrated detection of dopamine which is a neurotransmitter and they showed that graphene electrodes are able to discriminate dopamine even in the presence of interferants such as serotonin and ascorbic acid. Graphene electrodes have also been demonstrated for glucose detection. [48.e] Wang et al doped graphene with nitrogen with N atom composition varying between 0.11 to 1.35%. [48.f] This material showed a high electrocatalytic activity for reduction of $H_2O_2$ and a fast electron transfer kinetics for glucose oxidase, thus allowing glucose sensing at a low concentration of 0.01 mM. Even lower limits of 1 µM were achieved for glucose detection with metal decoration (Pt-Au) of graphene nanosheets.[48.g] Enzyme doping of graphene has also been shown to be effective for glucose biosensing.[48.h]

Ohno et al constructed a graphene field effect transistor (GFET) and modified it with immunoglobulin E (IgE) aptamers which enabled selective detection of IgE proteins.[48.j] The solution-gated GFET has been demonstrated also as a pH sensor with a response of 99 mV/pH.[48.n] Graphene doped with Pt nanoparticles was incorporated in an amperometric cholesterol sensor using the enzymes cholesterol oxidase and cholesterol estarese.[48.m] There is also significant interest in graphene as a material for high frequency FETs (see figure 7) [49] and spin devices.[50] An important aspect to be addressed in this regard is that the spin diffusion length is less than a few µm and the spin coherence only a few hundred ps. Another widely investigated application for graphene is in the construction of a supercapacitor [51] which is an energy storage device that attempts to combine the high power density of a capacitor with the high energy density of a battery. Conventional supercapacitors use activated carbon as the electrode material and more recently CNTs have also been investigated. The electrode material must have a high surface area, a porous network and high electrical conductivity. The device typically consists of a symmetric arrangement of two carbon electrodes backed by collector metals and separated by a porous membrane. The electrodes are soaked in either aqueous or organic electrolytes depending on the desired operating voltage. Stoller et al [51.a] mixed chemically modified graphene (CMG) with a binder material and shaped it into disk-like electrodes. Their devices showed 135 and 99 F/g for aqueous and organic electrolytes respectively. With a surface area of surface area 705 m2/g, their results translate into a specific surface capacitance of ~14 µF/cm2. Even higher specific capacitances of ~28 µF/cm2 using graphene nanosheets have been obtained by Du et al [51.b] without performance degradation over 500 cycles.

Graphene-polyanilyene composites have been shown to have a specific capacitance as high as 480 F/g. [51.c] A conductive tin oxide/graphene nanocomposite was investigated for supercapacitor application but only showed a 43.4 F/g capacitance.[51.f] ZnO also appears to yield rather low values for capacitance as a composite made with graphene.[51.g] In contrast, nanocomposites of graphene with manganese oxide showed capacitances in the range of 111-216 F/g.[51.h]

Qu et al used graphene as an electrocatalyst for oxygen reduction in fuel cells.[52] Nitrogen-doped graphene (N-graphene) was synthesized by chemical vapor deposition from a mixture of ammonia and methane in argon which showed superior electrocatalytic performance over undoped graphene as well as standard Pt electrodes. This, graphene can be a cheaper alternative to Pt in fuel cell applications. The CVD-grown graphene films are also ideal as transparent conducting electrodes in solar cell applications since they exhibit a low sheet resistance and high transmittance compared to indium tin oxide (ITO). In addition, the flexibility of the graphene films is ideal for organic photovoltaics since bending often results in cracks in the case of ITO.[53]

3.0 Graphene synthesis

The number of routes with which one can make graphene is enormous and ever increasing. In this section we review the more commonly applied routes. These for the most part have been popularized due to their scale-up potential. In addition, historically important and still relevant techniques such as the now infamous "scoth-tape" route are also included.

3.1 Chemical vapour deposition (CVD)

Chemical vapor deposition (CVD) is a well established technique and is often used to synthesize carbon-based nano-structures, such as carbon nanotubes and graphene. The underlying principle of CVD is to decompose a carbon feedstock (usually a hydrocarbon) with the help of heat in order to provide a source of carbon. The carbon can then rearrange to form sp$^2$ carbon species which is commonly accomplished over a catalyst. In the case of graphene, the most successful catalysts thus far are metals, however, unconventional routes using non-metallic substrates are beginning to emerge.

*3.1.1 Thermal CVD graphene formation over metals*

Graphene has been shown to be grown over a variety of transition metals, including nickel, [54] Iridium,[14, 55] Ruthenium [56] and copper [57] (see figure 8). In the case of Ni, small grain sizes and its high carbon solubility limit the size of the graphene sheets forming over a nickel surface.[58] Annealing the nickel substrates creates grains of 5 to 20 μm in size, which favor larger graphene sheet formation, but is still limited by the grain size.[58.b] Optimizing the thickness of a Ni film deposited on a substrate (e.g. Si/SiO$_2$) has also been shown to increase the size and the quality of the graphene sheets.[59] This is because the thickness of the Ni film can control the amount of carbon absorbed in the nickel foils over which the graphite crystals form upon precipitation.[58.a] Continuous ultrathin graphene films (1 to ~ 10 layers) can be prepared over polycrystalline Ni films (500 nm thick) by exposing the polycrystalline Ni film (at 900-1000 °C) to a highly diluted hydrocarbon flow at ambient pressure. Prior to the synthesis, the Ni films deposited on Si/SiO$_2$ substrates via e-beam evaporation can be thermally annealed in an Ar/H$_2$ atmosphere to maximize the Ni grain sizes. [58.b] The use of H$_2$ in an annealing pre-synthesis step leads to smoother surfaces, which leads to fewer and more uniform graphene layers. It is also believed that hydrogen annealing removes impurities (such as S and P) which can cause local variations of carbon solubility in the metal substrate.[54.b] In addition, it is argued, H can remove defects in carbon at elevated temperatures.[60]

Highly crystalline large-area graphene over polycrystalline Ni surfaces can also be achieved at low pressures (10$^{-3}$ Torr) with temperatures below 1000 °C. The technique requires short growth times and rapid quenching to yield highly crystalline few-layer graphene. The choice of quenching rate is related to the solubility of C in Ni and the kinetics of C segregation. With slow cooling rates, the C has sufficient time to diffuse into the bulk and no segregation is found on the surface, because Ni has a high carbon solubility. On the other hand, with high cooling rates C atoms segregate out of Ni very rapidly leading to less crystalline, defective graphitic structures. Hence the cooling rate is crucial, and only at moderate cooling rates do C atoms segregate and form highly crystalline graphene.[54.b] Wrinkles are often observed in the as-grown few layer graphene sheets, as shown in figure 8(e). The wrinkles form independently of the grain boundaries and two processes are suggested for their formation : *i.* nucleation of defect lines on step edges between Ni terraces and *ii.* thermal-stress-induced formation of wrinkles around step edges and defect lines (see figure 9).[61] In order to minimize defects in the graphene layer introduced at the boundaries of the grains of polycrystalline Ni, the use of single crystal Ni films are being explored. A comparison study between growth of graphene on single crystal Ni (111) and polycrystalline Ni by Zhang et al.[62] show that it is possible to have more than 90 % coverage of monolayer/bilayer graphene on single crystal Ni, while the coverage is only about 72% for polycrystalline Ni. They argue that this is due to the atomically smooth surface and grain-boundary-free structure of the single crystal Ni. Growth of graphene layer on single crystal Ni (111) at low temperatures has also been reported.[63] The Ni substrate was exposed to ethylene at $10^{-5}$ Torr in an ultrahigh vacuum (UHV) chamber at 460 °C which causes the formation of a nickel carbide layer. Keeping the sample at 460 °C in vacuum, the nucleation and growth of graphene on pure Ni is observed. This is due to the segregation of carbon from the bulk to replace the Ni atoms in the $Ni_2C$. The formation of carbide layer is reported when the growth temperature is less than or equal to 460 °C.

Graphene can also be grown on Ir (111) via CVD in ultrahigh vacuum (UHV). Coraux et al sputtered Ir on a silicon substrate and then annealed it at high temperature to yield clean Ir (111) with terraces extending over several hundreds of nanometers. Ethane was used as carbon precursor.[64] They found in the case of low graphene coverage that the graphene is only located at the substrate step edges. In addition, the graphene frequently spans both sides of the step edge with the largest fraction of graphene residing on the lower terrace. Moreover, they showed that with their CVD process it is possible to tune the graphene coverage up to 100 % with a very high structural quality when implementing temperatures above 847 °C and carefully controlling the ethane dose.[64] In another study, large areas of single or few layer graphene were grown using low-pressure chemical vapor deposition (CVD) with ethylene as the feedstock on hot Ir (111) surfaces. The graphene sheets were shown to lie like a blanket on the substrate and had bends over step edges.[14] Epitaxial graphene has also been grown over Ru (0001) surfaces. Sutter et al controlled the cooling rate in order to make use of the temperature-dependent solubility of interstitial carbon similar to that discussed above for Ni substrates.[56.b] They were able to achieve controlled layer-by-layer growth which formed large lens-shaped islands of graphene on the Ru (0001) surface. Copper has been shown to be a more promising support for large area graphene synthesis via CVD. Unlike Ni, copper has a very low C solubility and large grain size formation after annealing which enables one to grow graphene sheets over large areas.[57.a] The dependence of the Cu film thickness, substrate type and reaction time have been extensively investigated by Ismach et al.[59] They showed the continuity of the metal film on the surface and hence the graphene film depends on its thickness, the metal-dielectric wetting properties, and the heating temperature and time. Indeed, during the heat-up period the Cu film dewets and is dependent on the film thickness. The dewetting process causes discontinuity of the graphene layer in metal-free areas. Hence, the film should be thick enough to minimize metal-free regions so as to maximize the graphene growth area. The dewetting process also causes wrinkles in the graphene sheets. This induces stress on the graphene sheet, possibly rupturing the sheet which can lead to subsequent new graphene sheet nucleation. Li et al.[57.b] studied the effect of growth temperature, methane flow rate, and methane partial pressure on graphene sheet formation over polycrystalline Cu. They obtained graphene islands, mostly with a star like shape, and uniform density across the Cu foil (see figure 8). They were able to show a direct proportionality between the graphene nuclei (domain size) with methane flow rate, and methane partial pressure in the CVD process. The graphene nuclei are inversely proportional to the growth temperature. By optimizing the parameters they claim that full surface coverage of graphene is possible. Similarly, hexagonal single crystal islands of few layer graphene with pyramid-like shape were grown on Cu foils.[65] They found that the islands follow a preferential direction and they may be related to step edges or folds. They showed that the domain size as well as number of layers in the islands is directly proportional with the growth time.

*3.1.2 Thermal CVD graphene formation over dielectrics*

One of the challenges currently facing graphene based nano-electronics is obtaining defect free and clean graphene on suitable substrates. The substrates need to be insulating with a high dielectric constant, K, to minimize current leakage when used as a gate. However, the fabrication of graphene devices with an atomically uniform gate dielectric which can provide a uniform electric field over the active graphene area remains unresolved. Apart from epitaxial growth on SiC, there are very few other routes directly forming graphene on the surface of a dielectric. Moreover, the strong coupling between the atomic geometries and electronic structure of SiC growth surface and graphene limits its development in electronics. Hence there is a need for alternative routes. One possibility might be through the use of Plasma enhanced CVD (PECVD) in which the formation of free-standing Carbon nanosheets and planar graphite films with a few graphene layers have been demonstrated.[66] PECVD is a promising candidate for mass production of 2D graphene nanosheets, because of its simplicity and compatibility with traditional semiconductor processes. Moreover it offers a high degree of control.[67] Often a traditional catalyst is used as the substrate for their production; however, various reports show pretty much any material can be used, including Si wafers with an oxide surface layer.[68] Dato et al [69] showed that one can even form graphene sheets without a substrate, for example, directly in the gas phase. Conventional thermal CVD can also yield graphene over non-metallic catalysts. Recently, thermal CVD at temperatures between 1350 and 1650 °C using propane as the carbon feedstock was shown to form graphitic layers over sapphire.[70] Another group demonstrated graphene nano flakes can form on a variety of oxides using thermal CVD.[71] Moreover they have established it can be achieved with temperatures below 500°C when using acetylene as the feedstock.[72] Their experimental findings also show that the reaction stops once the oxide supports (nano-crystals) are encapsulated with graphitic layers. In addition, the layers terminate at step sites, which suggests that step sites on oxides may be important for graphene nucleation and growth. Preliminary theoretical investigations using MgO (100) surfaces with step sites as a model system for oxides show acetylene does not easily adsorb on the surface but does so more readily at steps sites. Moreover, acetylene dissociation on the surface is shown to be endothermic whilst at a step site the process is exothermic. Complimentary diffusion studies show a diffusion barrier of ca. 0.38 eV for C and 0.2 eV for H over an oxygen atom on the MgO (100) surface. This points to C remaining on the surface whilst H simply flies away. The authors propose a graphene formation mechanism involving four distinct sub-processes; the adsortion of the carbon feedstock molecules on the substrate surface, dissociation of hydrogen from the precursor, surface diffusion, and addition of carbon atoms to the network.[72.b] These sub-processes are shown in figure 10. A further route takes advantage of the efficiency with which metals form graphene in thermal CVD. In essence, a Cu sacrificial layer on the substrate dewets and evaporates during synthesis. [59] In this manner one can form graphene pseudo-directly on a dielectric. In a similar pseudo-direct route, graphene-like thin films can be grown in the vicinity of an edge of a lithography-fabricated pattern on a $SiO_2$ substrate.[73]

## 3.2 Graphene growth via carbon segregation

An alternative to the CVD process but which is still based on the precipitation of carbon from metals are thin film carbon segregation routes (see figure 11).[74] This method has advantages because the additional involvement of surface adsorption of extraneous organic molecules, catalytic decomposition, and dissolution of carbon species into bulk metal is not necessary.[74.c] In one example, graphite, as a carbon source, was dissolved in molten nickel at 1500 °C for 16 hours in a vacuum of $10^{-6}$ torr. Single layer graphene was then grown by controlling the cooling rate such that the dissolved carbon atoms are able to segregate from the nickel surface forming graphene.[74.a] In another system, a Ni thin film was epitaxially grown on highly oriented pyrolytic graphite (HOPG). Under optimized high vacuum and temperature conditions (650 °C for 18 h), carbon atoms start to segregate from the HOPG through the Ni film to form a single layer graphene on the free face of the Ni thin film.[74.b] Using this route they were able to grow almost defect-free single layer graphene with the size of 2 cm × 2 cm. In another study, wafer size graphene growth was recently achieved.[74.c] Here, commercial metals containing stable carbon impurity as carbon source were evaporated directly onto $SiO_2$/Si wafers. Few layer graphene (1-3 layers) were grown by annealing at 1100 °C in a vacuum of $4 \times 10^{-3}$ Pa. As a result of the vacuum annealing process, carbon atoms diffuse toward the surface and then segregate forming graphene. Later, Zhang et al developed a route in order to grow nitrogen-doped graphene based on this segregation mechanism.[74.d]

## 3.3 Epitaxial Graphene Growth on SiC

A widely used technique to synthesize graphene is the thermal decomposition of hexagonal α-SiC (6H-SiC and 4H-SiC). It has the advantage that it is very clean because the epitaxially matching support crystal provides the carbon itself and no metal is involved. The technique dates back to the early 1960s when Badami through X-ray scattering studies found graphite on SiC after heating it to 2150 °C in UHV.[75] He proposed the graphite formed through preferential surface Si out-diffusion. A decade later, van Bommel and coworkers found monolayers of graphite on SiC after annealing at 800 °C and enhanced graphitization around 1500 °C. They also identified that the observed variation of the crystallinity of the graphene layers depended on the surface termination, namely, the Si-terminated (0001) face and the C-terminated (000 ) face. They found epitaxial alignment of the graphene layer on C-terminated (000 ) faces to be rotated 30° with respect to the Si-face unit cell.[76] Later in the 1980s and 1990s, Muehlhoff et al., Forbeaux et al., and Charrier et al. conducted more detailed studies on the graphitization process which confirmed the results of van Bommel.[77] The parallel publication of the electrical response of graphene in 2004 by Novoselov et al. and Berger et al. (who used graphene grown from SiC) provided new impetus to optimize the growth conditions of graphene on SiC [2.a, 78] (see figure 12).

3.3.1 Si sublimation at high temperatures

The synthesis of graphene from Si sublimation of SiC starts with a surface pre-treatment to remove oxides and to clean and smoothen the surface. Several different methods have been reported. The importance of this initial step is not clear [79] because the synthesis temperatures are much higher, which in itself leads to significant surface degradation. The most popular method for removing polishing scratches on the surface of high quality SiC wafers is via hydrogen etching in which after an initial pre-cleaning step by ultrasonicating in acetone or ethanol, the samples are heated in a $H_2$/Ar (5%/95%) atmosphere to temperatures of around 1600 °C for a period of ca. 60 min. Subsequent cooling rates should be slow (around 50 °C/min) to avoid Si crystallization on the surface. This procedure yields a flat surface with atomic steps whose distance is dependent on the angular deviation of the cut with respect to the crystallographic direction (see figure 13). Any remaining oxide layers can be removed by annealing in UHV at temperatures of 950 °C–1100 °C for several minutes. To avoid excessive Si sublimation, a Si flux (~$10^{14}/cm^2$ s) helps maintain the surface stoichiometry. Here, the temperature must be kept between 850 °C and 900 °C to prevent either the formation of a crystalline Si surface layer or etching.[79] The majority of the graphene growth is obtained on the Si face of SiC. In the course of heating SiC to the graphene formation temperature, the surface undergoes several phase changes. Finally, at temperatures of 1250 °C–1350 °C the $(6\sqrt{3} \times 6\sqrt{3})R30$ reconstruction is reached which is Si depleted.[77.b] It is from this surface reconstruction that graphene grows. A difficulty with this technique is the occurrence of surface roughening due to Si loss which causes the graphene to grow in flakes. A further complication is that there exist interactions between the first graphene layer and the substrate which influences the electrical characteristics of graphene such that the covalent bonds disturb the linear π bands. One can decouple the graphene from the substrate. Riedl et al. intercalated hydrogen between the graphene and the substrate. This saturates the topmost Si layer and thus frees the (graphene) π bands.[80] Another approach, explored by Oida et al., is to introduce an oxide buffer layer.[81] This can be done by heating the sample to 250 °C in 1 atm $O_2$ for 5 s. At higher temperatures the graphene is etched. An alternative approach is to reduce the process temperature by introducing a partial pressure of HCl in the Ar atmosphere assist in the removal of the first Si layers.[82]

The mechanism of the nucleation and growth of graphene from SiC has, in general, been well explored. Hannon and Tromp observed the development of deep pits in the substrate during annealing in vacuum in an in situ study.[83] The pits form because domains from the buffer layer pin decomposing surface steps. Graphene has been observed to nucleate in pits where the step density is high. Hupalo et al. found that the Si desorption from steps is the main controlling process for graphene nucleation. [84] They also showed that different desorption rates for different step directions and that faster heating rates could suppress these different speeds leading to significantly larger graphene sheets. TEM studies by Norimatsu and Kusunoki and Robinson et al. revealed the preferential nucleation site at step like facets on the SiC surface.[85] Taking advantage from this, Sprinkle et al. used the facets as templates for bottom-up growth of graphene ribbons for top-gated transistors.[86]

4.2.2 Variations

Al-Temimy and coworkers demonstrated the growth of graphene on SiC at temperatures of 950 °C via molecular beam epitaxy.[87] This method reduces the degeneration of the SiC surface quality and thus larger graphene sheet sizes are possible. Interestingly, the grown graphene does not show the 30° rotation as found with conventionally grown graphene on SiC. Aristov et al. demonstrated the feasibility of graphene synthesis on cheap, commercially available cubic β-SiC/Si wafers of > 300 mm in diameter. This route is compatible with industrial mass production. ARPES studies revealed that the interaction of the graphene with the substrate is very weak and is much lower than that reported for the graphene–α-SiC system. This preserves the intrinsic properties of graphene further making the technique attractive.[88] Low synthesis temperatures of around 750 °C for graphene on SiC were reported by Juang and coworkers.[89] The ploy here is to cover the SiC with a Ni thin film that dissolves the carbon from the substrate surface and then precipitates on its free surface upon cooling. Hofrichter et al. used the same principle to produce graphene on $SiO_2$ through the preparation of a layer stack of amorphous SiC and Ni thin films at 1100 °C.[90] Another approach that could yield graphene is the carbo-thermal reduction of $SiO_2$ to SiC and subsequent graphitization as demonstrated by Bachmatiuk and coworkers.[91] They investigated the catalytic potential of $SiO_2$ nanoparticles for carbon nanostructure growth in a CVD process and found that the particles that nucleated stacked-cup nanostructures were converted to SiC. The technique is now being developed for graphene and shows promise (see figure 14).

3.4 Exfoliation Routes

Due to ease of production and low cost, the exfoliation of graphite is currently the most popular route to prepare graphene. The more common exfoliation routes include mechanical exfoliation,[2.a] ultrasound treatment in solution [92] and intercalation steps.[93] Nevertheless, the quality of the obtained graphene prepared by exfoliation is not always sufficient. One of the biggest problems is that often the homogeneity of the number of graphene layers is poor. Moreover, the purity of the produced contains remains from the exfoliating agents.

*3.4.1. Mechanical exfoliation of graphite*

Mechanical exfoliation of graphite is also known as the ''Scotch tape'' or peel-off method. The method entails the repeated peeling of small islands of graphite (usually HOPG – highly oriented pyrolitic graphite or natural graphite) using cellophane tape. During the preparation of graphene a surface of a graphite crystal is rubbed against another surface *viz.* adhesive tape. The number of layers forming a flake can be controlled to a limited degree through the number of repeated peeling steps. The obtained flakes can be then transferred to different surfaces like a Si wafer for further studies or device fabrication. This method is highly reliable and allows the preparation of few-layer graphene films up to 10 µm in size. However mechanical exfoliation of graphite yields small samples of graphene that can be used mostly only for fundamental studies.[2.a] Another disadvantage is that the samples can contain impurities from the adhesive tape.

*3.4.2. Ultrasound treatments of graphite in liquid solutions*

The ultrasound route involves the exposure of graphite powders to ultrasound in organic solvents, for example, dimethylformamide (DMF) or N-Methyl-2-pyrrolidone (NMP).[92, 94] The choice of organic solvent is related to the solvent-graphene interaction, so that the surface energy of the solvent matches that from graphene.[92] Usually ~ 0.1 mg/ml solution of graphite in a relevant solvent is sonicated in a low power sonicator for a short time ~ 30 minutes. An alternative ultrasonication route entails the use of a water-surfactant solution, sodium dodecylbenzene sulfonate (SDBS).[95] Here, the graphite powder is dispersed in a SDBS water solution and then placed in a low power sonication bath. The choice of ultrasound intensity is important so as to minimize the structural damage to the graphene. After sonication, the obtained solutions are usually centrifuged at ~ 500 rpm for around 90 minutes. The upper part of the resultant supernatant contains single layers of graphene floating in the solution. The supernatant can be transferred using a pipette and then dropped on the surface of choice for further studies e.g. on a silicon wafer or a TEM grid.[92, 95] Exfoliation by ultrasonication usually results in a monolayer graphene yield of around 1 wt. % but can be as high as 12 wt. %.[92]

*3.4.3. Intercalation of graphite*

The basis behind the intercalation route is to introduce species in between graphene layers which can be treated in a certain manner such that they expand forcing the graphene layers to split apart from each other. This is often done by introducing sulphuric and nitric acids between the graphene layers and then quickly heating the samples in an air atmosphere at ~ 1000 °C [93] The rapid evaporation of the intercalates yields exfoliated graphite flakes with a size around some tens of nanometers. The samples contain large amounts of single graphene sheets which can be separated from the larger flakes via ultacentrifugation.[96] Oleum and tetrabutylamonium (TBA) have also been shown to be good intercalants. [96.b] In these cases the samples are sonicated in the surfactant solution. This leads to a suspension consisting of few layer and single layer graphene forming. Again, the separation of the different fractions can be obtained through centrifugation. [96.b]

*3.4.4. Reduction of graphene oxide*

The reduction of graphene oxide solutions by different reducing agents is another route to form graphene. Commonly used reducing agents are: sodium borohydrate,[95] ascorbic acid,[98] hydrazine [99] and hydroquinone.[100] Graphene oxide itself consists of small graphene flakes which are chemically functionalized. The functionalization of graphene oxide renders it hydrophilic allowing it to be dispersed in water based solutions. The starting material to produce graphene oxide is graphite oxide which can be produced by different routes, namely the Hummers [101], Brodie [102] or Staudenmaier [103] methods. The Hummers method involves soaking the graphite powder in a solution of sulphuric acid and potassium permanganate.[101] The Brodie and Staudenmaier methods use potassium chlorate and nitric acid mixtures as an oxidizing agent. One can also prepare graphite oxide using strong acids like nitric, sulphuric or perchloric acid.[104] Stirring or sonicating the graphite oxide solutions is then performed to obtain single layers of graphene oxide. Single layers of graphene oxide platelets can be separated from larger particles via centrifugation. Finally, the graphene oxide is treated with reducing agents, by thermal treatment [105] or electrochemically [106] in order to remove surface functional groups and yield graphene.

3.4 Functionalized graphene

One of the key driving forces to fabricate graphene with atomic precision in a reliable and economical manner is the desire to open a band gap. Apart from the formation of a band gap by introducing confinement effects, namely, nano ribbons, one can open an energy gap through functionalization. The prominent functionalization routes are briefly reviewed in this sub-section. This is followed by an introduction to nanoribbon formation.

*3.4.1 Graphane*

Graphane is the hydrogenated equivalent of graphene. It is composed of carbon and hydrogen atoms forming a two dimensional hydrocarbon and is written as $(CH)_n$. The hydrogenation changes the $sp^2$ structure of graphene to a $sp^3$–like hybridization. Graphane can be synthesised using a stream of hydrogen atoms,[107] reactive ball milling between anthracite coal and cyclohexane,[108] exchanging the fluorine in fluorinated graphite by hydrogen [109] or by dissolved metal reduction in liquid ammonia.[110]

*3.4.2. Fluorinated graphene*

Fluorinated graphene (FG) is graphene functionalized with fluorine. It has a wide band gap, and possesses insulating properties. Its elasticity and mechanical strength is similar to graphene.[111] There are different methods to produce fluorinated graphene (FG), namely the extraction of single layers of FG from commercially available fluorinated graphite,[2.b] the exposure of graphene to fluorine gas at ~ 500 °C,[112] placing graphene in a fluorine-based plasma,[113] or exposure to xenon difluoride.[114]

*3.4.3. Graphene in combination with boron and nitrogen*

The strong lattice correlation between the carbon atoms in graphene sheets and the B and N in BN sheets, opens new opportunities for growing high quality graphene layers over BN substrates,[115] BCN graphitic films (thickness from 100nm to a few micrometres). BCN graphitic films can be grown via CVD using B-, C- and N-containing precursors [114] which leads to graphene formation with B and N also incorporated. This doping is desirable in order to tune their bandgaps.[23, 116] These doped graphene structures also have interesting mechanical properties.[117]

*3.5 Graphene Nanoribbons*

Graphitic (graphite or graphene) nanoribbons (GNR) are defined as one-dimensional sp$^2$ hybridized carbon crystals with boundaries that expose non-three coordinated carbon atoms, and possess a large aspect ratio. As a result of size confinement these structures can have a band gap and are hence attractive as building blocks for molecular electronics. Graphene nanoribbons have been produced by both top-down and bottom-up approaches. The top-down fabrication of GNR by opening (unzipping) carbon nanotubes [118] and etching graphene sheets [96.b, 119] has been demonstrated. The unzipping of carbon nanotubes to yield GNR can be achieved via intercalation/exfoliation. In this route, the tubes (MWNT) are intercalated with lithium (in liquid ammonia) followed by acid and thermal exfoliation treatments.[118.a] An alternative two-step MWCNT unzipping process has been developed by Jiao et al.[118.b] Pristine MWNTs synthesized by arc discharge were first calcined in air at 500 °C in order to remove impurities and etch/oxidize MWNTs at defect sites. The nanotubes were then dispersed in 1,2-dichloroethane (DCE) organic solution of poly(m-phenylenevinylene-co-2,5-dioctoxy-p-phenylenevinylene) (PmPV) by sonication. During this process the calcined nanotubes unzip into nanoribbons. Seperation of the ribbons via ultracentrifuge yields up to 60 % GNR with nearly atomically smooth edges. A similar approach has been introduced by Li et al,[96.b] where they used an easily available and abundant graphite material (Grafguard 160-50N) instead of MWCNT in order to fabricate GNR. Kim et al.[118.c] used current-induced electrical breakdown of MWNT to fabricate graphene nano-ribbons. Careful selection of the proper voltage bias causes only part of the MWNT outer wall to sever and a precursor GNR is created which clings to the remaining MWNT inner core. Then, a movable electrode is used to systematically slide the GNR from the MWNT inner core. An advantage for this technique is that the length of the GNR is controllable in the sliding step. Opening MWCNT can also be achieved through oxidation.[118.d] In this work, MWCNTs were opened via permanganate oxidation in acid which leads to the unzipping of the nanotubes along their axis. High yields nearing 100% are possible and they were shown to be suitable to build field effect transistors with extremely high electrical conductivities.[118.e] In another approach, GNR can be produced by removing longitudinal strips of carbon atoms from the side walls of MWNTs which are partially embedded in a polymer film. Etching the exposed carbon using an Ar leaves only the embedded material, *viz*, nano-ribbons. The technique allows one to produce single-, bi- and multilayer GNRs with different widths depending on the diameter and number of layers of the starting tubes and the etching time.[118.f] Bai et al [118.g] demonstrated a similar method using an oxygen plasma. The use of metallic catalyst particles to "cut" graphite/graphene is also possible through hydrogenation or oxidation processes. In the hydrogenation process, metal nano-particles are exposed hydrogen at elevated temperatures. The particles etch the graphene releasing $CH_4$ in the process (see figure 15).[119.a,b, 120] The etching of graphite using different metals and oxides at elevated temperatures in different atmospheres, such as oxygen,[119.d,e] carbon dioxide and water [119.f] is also possible. However, there remain limitations and drawbacks with these top-down approaches such as limited amounts of produced ribbons, scale-up difficulties, accurate and reproducible edge control, production time and high cost, to name a few.[121] More recently, a bottom-up approach has been demonstrated for fabricating GNR using 10,10'-dibromo-9,9'-bianthryl as precursor monomers. This may be a more promising approach.[122] During a first thermal activation step at a temperature in the range (200 – 250 °C), the biradical species diffuse across the surface of a support and undergo radical addition reactions to form linear polymer chains as imprinted by the specific chemical functionality pattern of the monomers. In a second thermal activation, which is performed in a temperature range between 400 and 440 °C, step surface-assisted cyclodehydrogenation establishes an extended fully aromatic system. The presented approach is compatible with the used CMOS technology, since the temperature range for it is low compared to other techniques. The formation process for these bottom up grown nanoribbons is illustrated in figure 16. Another low temperature route for graphene nanoflake synthesis directly on dielectrics is via thermal CVD. [72.a]

3.6 Defects in Graphene

Defects in graphene, point defects as well as extended defects, as has already been discussed, have a great impact on its electronic and mechanical properties. [123] Hence, apart from the need to fabricate graphene with atomic precision, it is also imperative that we are able to do so in a defect-free manner, unless, of course, we desire the defect and can achieve this, again, with atomic precision. The difficulties of corrugation are often incurred in suspended graphene as reported by Meyer et al.[124] Recent advances in transmission electron microscopy, *i.e.* aberration corrected lenses, allow one to image point defects in graphene and to follow the reconstruction process.[125] Stone-Wales defects have been observed to form and relax to a perfect lattice again and vacancies to reconstruct to pentagon-heptagon structures.[126] Also, the structure and stability of graphene's edges have been investigated. Liu et al. found that two adjacent layers tend join and form a folded edge to minimize the number of dangling bonds.[127] In single layers, the zigzag edge has been found to be the most stable configuration.[128] Rotational stacking faults have found to effectively decouple the layers electronically and let multilayer graphene behave like single graphene sheets. Warner et al. showed the reconstruction of the packing structure of up to six rotated sheets. [129] In addition, the large area synthesis of graphene with many growth nucleation centres introduces grains with relative rotation.[130] These grains stitch together via pentagon-heptagon pairs and are in essence defects which affect its conductivity. [3] In short, there remains a need to produce defect free graphene in large quantities and in a reproducible manner.

4.0 Conclusion

Since the seminal works of Geim and Novoselov [2] there has been an explosion in interest in graphene. There is no doubt that this wonderful material will have an impact on our everyday lives sometime soon. For example, Vorbeck, a materials company based in the USA claims to have developed lithium ion batteries with a special graphene material that can recharge in 10 minutes. The use of graphene to form lightweight, strong and electrically tunable composites will also emerge in many applications. However, in order for graphene to be a serious contender beyond silicon technology, enormous strides on several fronts in graphene synthesis are required. Key aspects to be addressed are the need for grain and grain boundary-free graphene, producible on large areas and at suitable temperatures. Moreover, this must be accomplished in a manner that is economically viable. As has been highlighted in this review, current graphene synthesis has yet to deliver on these requirements. It is for this reason the most successful devices thus far have been fabricated using HOPG transfer. Future synthesis research will most probably see the delivery of the above requirements. The challenge is tremendous with regards to the synthesis of ribbons with defined edges and widths, or functionalized derivatives with reproducible and definable properties. There is hope though. The entry of graphene comes on the back of fullerenes and carbon nanotubes. Many of those involved in their synthesis are turning to graphene bringing valuable and relevant knowledge with them. Furthermore, researchers with other backgrounds, for example, in SiC are also actively contributing useful knowledge. Moreover, the atomically precise bottom up growth of graphene nanoribbons by interlinking intermediate structures derived from polymers is an exciting and important breakthrough [122]. At the very least, the current achievements in graphene synthesis along with the shear momentum behind graphene research as whole promise some mouthwatering developments in the future.

Acknowledgements
MHR thanks the EU (ECEMP) and the Freistaat Sachsen, II the DAAD (A/07/80841), FB the DFG (RU 1540/8-1), AB and CGR the Alexander von Humboldt Foundation and the BMBF, MM and GC acknowledge support by the World Class University program through the National Research Foundation of Korea funded by the Ministry of Education, Science and Technology under Project R31-2008-000-10100-0. We are grateful to Sandeep M. Gorontla for help in the production of the TOC figure.

Received: ((will be filled in by the editorial staff))
Revised: ((will be filled in by the editorial staff))
Published online: ((will be filled in by the editorial staff))

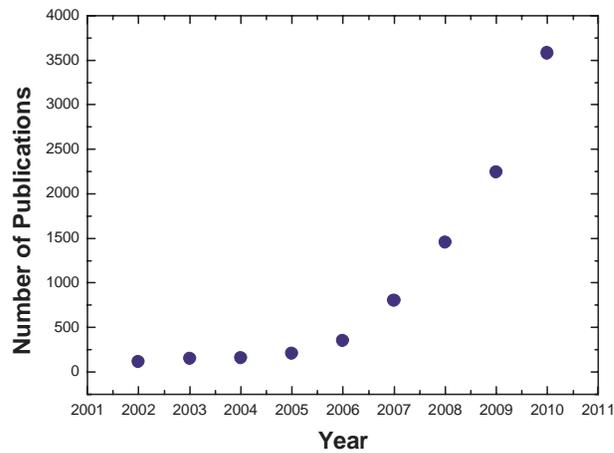

**Figure 1.** Annual publications under the subject area Graphene from 2002 to 2010. Data collected from ISI web of Science.

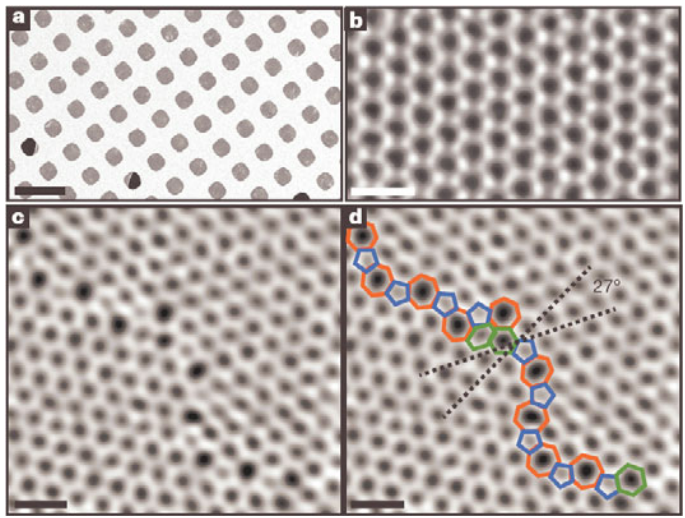

**Figure 2.** Atomic-resolution ADF-STEM images of graphene crystals. a, Scanning electron microscope image of graphene transferred onto a TEM grid with over 90% coverage using novel, high-yield methods. Scale bar, 5 mm. b, ADF-STEM image showing the defect-free hexagonal lattice inside a graphene grain. c, Two grains (bottom left, top right) intersect with a 27u relative rotation. An aperiodic line of defects stitches the two grains together. d, The image from c with the pentagons (blue), heptagons (red) and distorted hexagons (green) of the grain boundary outlined. b–d were low-pass-filtered to remove noise; scale bars, 5A˚. Reproduced with permission from [3] Copyright 2011, NPG.

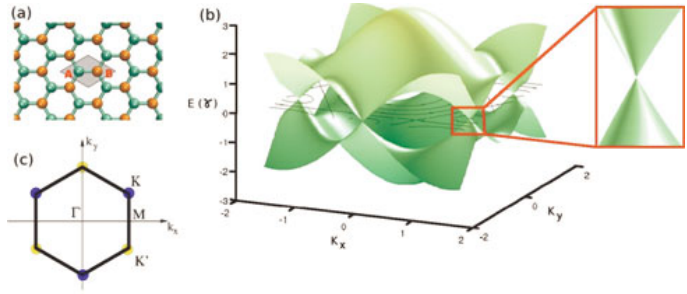

**Figure 3.** (a) Honeycomb lattice of graphene. The shadowed area delineates the unit cell of graphene with its two non-equivalent atoms labeled by A and B. (b) Band energy dispersion obtained via tight binding approximation. The inset highlights the conical-shape dispersion around the charge neutrality point. (c) First Brillouin zone.

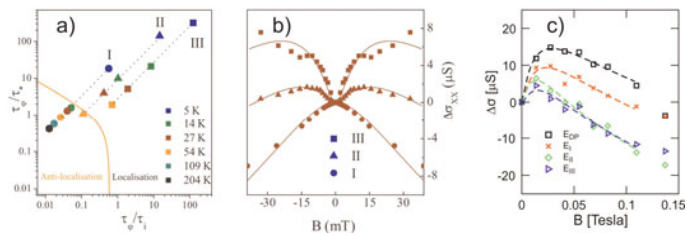

**Figure 4** Weak localization and weak antilocalization in graphene from experiment (a,b) and theory (c). Reproduced with permission from [15.c]. Copyright 2006, APS. Reproduced with permission from [13.a]. Copyright 1980, Physical Society of Japan. a) Temperature dependence of WL and WAL signatures for varying Fermi level positions Reproduced with permission from [15.c]. Copyright 2008, APS. b) and c) Magnetoconductance responses of experimental sample Reproduced with permission from [15.c]. Copyright 2008, APS and numerical simulation. Reproduced with permission [13.a]. Copyright 1980, Physical Society of Japan, respectively.

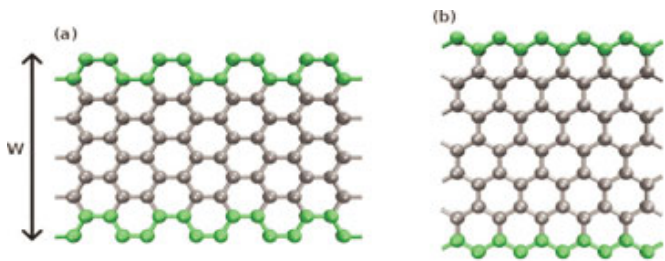

**Figure 5** Atomic structure of an (a) armchair- and a (b) zigzag- edge graphene nanoribbon. Green colour atoms delineate the respective edge-shape and W denotes the width of the ribbon.

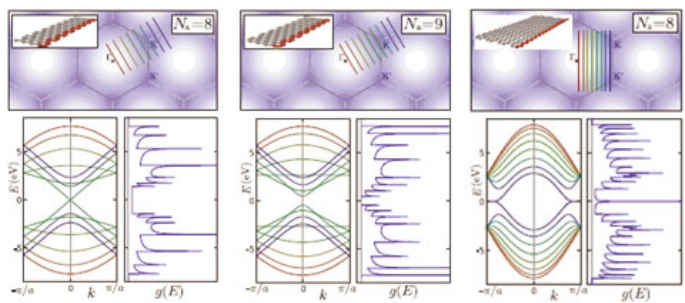

**Figure 6** (Top panels) Zone-folding diagram for three different graphene nanoribbons: (left) AGNR(8), (middle) AGNR(9) and (right) ZGNR(8). Their respective energy band structures and density of states curves are displayed on the lower panels.

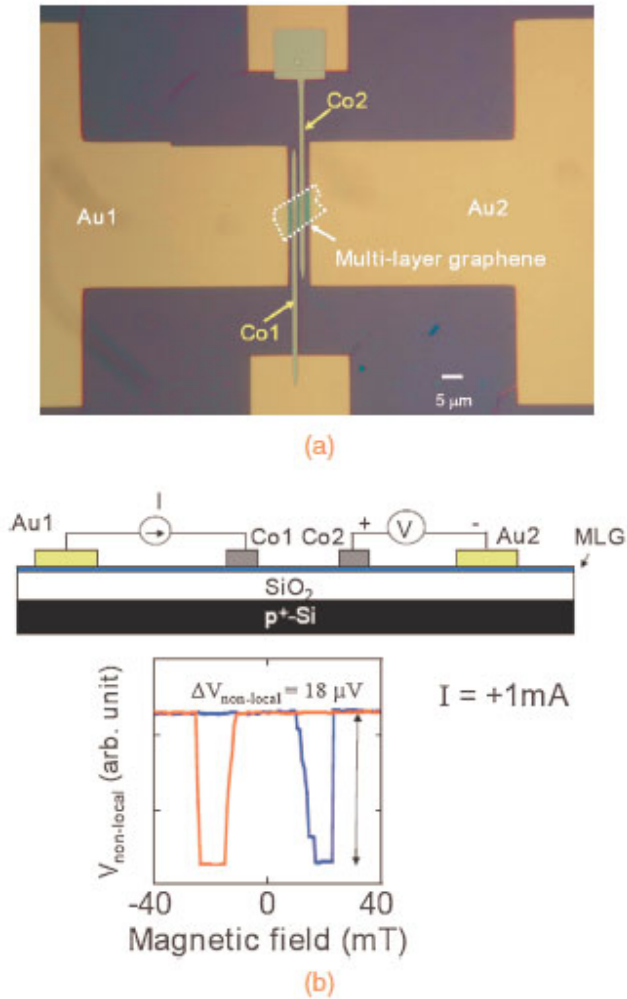

**Figure 7** (a) An optical microscopic image of an MLG spin valve. (b) A schematic view of the measurement circuit of spin injection into the MLG spin valves and a typical spin injection signal at an injection electric current of +1 mA. The electric current is injected from the Co1 electrode to the Au1 electrode (defined as positively biased in this study).    Vnon local is defined as the difference between the spin injection signal intensities for parallel and antiparallel magnetization alignments.
Reproduced with permission from [49] Copyright 2009, JSAP.

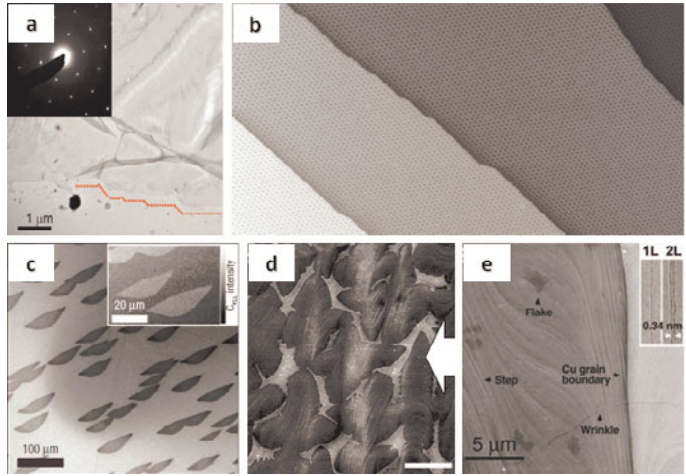

**Figure 8:** (a) TEM image of graphene grown on Ni substrate. (Inset) SAED pattern of the graphene film. Reproduced with permission from [56.b] Copyright 2008, AIP. (b) STM image of graphene grown on Ir(111). Reproduced with permission from [66] Copyright 2009, IOP. (c) UHV-SEM image of graphene islands grown on Ru(0001) surface. (Inset) UHV scanning Auger microscopy image, obtained on the sample. Reproduced with permission from [58.b] Copyright 2008, NPG. (d) SEM images of graphene grown on Cu substrates. Reproduced with permission from [59.b] Copyright 2010, ACS. (e) HR-SEM image of graphene grown on Cu foils. (Inset) shows TEM images of folded graphene edges. Reproduced with permission from [59.a] Copyright 2009, AAAS.

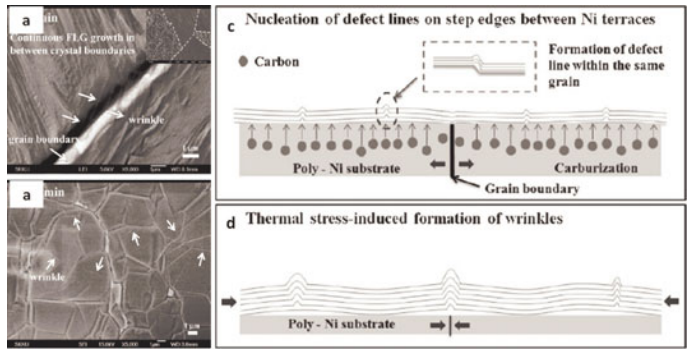

**Figure 9:** FESEM images of synthesized graphene layers on poly-Ni substrate grown for (a) 5 minutes showing wrinkles crossing over the grain boundary (b) 15 minutes with wrinkles on the inner grains. White arrows indicate wrinkles, 1 µm. Schematic diagrams of wrinkle formation: Wrinkles are generated because of (c) nucleation of defect lines on step edges between Ni terraces, (d) thermal-stress-induced formation of wrinkles around step edges and defect lines. Reproduced with permission from [61] Copyright 1993, Elsevier.

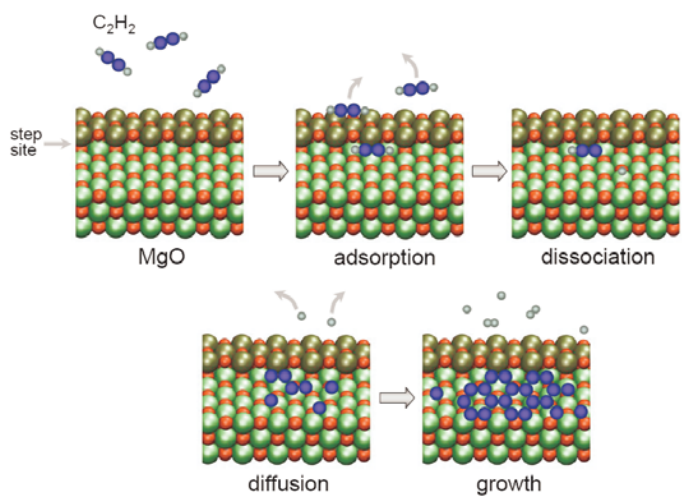

**Figure 10:** Proposed growth processes for graphene at step sites on oxides. $C_2H_2$ is preferentially adsorbed at step sites, where it dissociates, thereafter H diffuses away and finally C addition leads to graphene growth. Blue = carbon, gray = hydrogen, red = oxygen, green/gold = Mg).

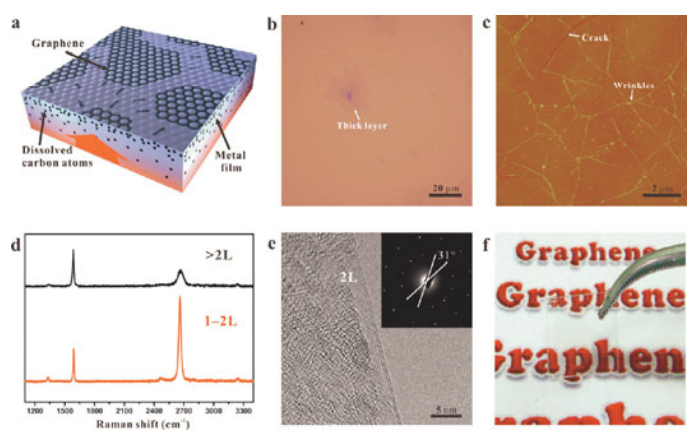

**Figure 11.** Segregation of graphene from bulk metal. (a) Schematic illustration of segregation technique, which demonstrates the surface accumulation of buried carbon atoms in bulk metal and formation of graphene at high temperature. (b-f) Characteristics of a graphene film segregated from polycrystalline Ni. (b) Optical microscope image of segregated graphene (shortly, s-graphene) transferred on 300 nm SiO2/Si substrate. (c) Tapping mode AFM height image showing the existence of wrinkles and crack in monolayer or bilayer regions. (d) Raman spectra of 1-2 L (red) and >2 L (black) graphene. The excitation wavelength was 632.8 nm. (e) TEM image and selected-area electron diffraction pattern (SAED) revealing the nice crystallinity and disordered A-B stacking structure of bilayer s-graphene. (f) Photograph of graphene film transferred to quartz substrate. Reproduced with permission from [74.c] Copyright 2011, ACS.

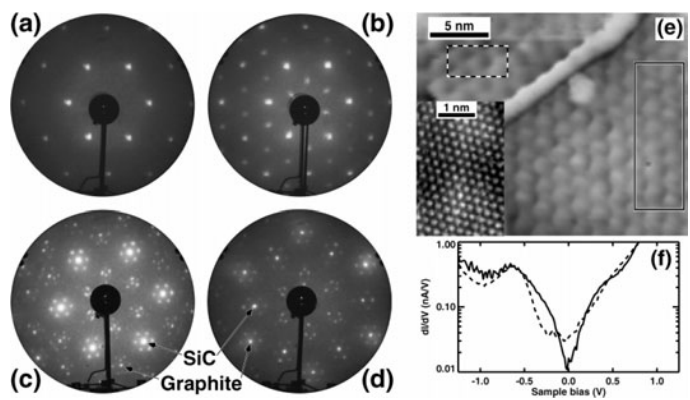

**Figure 12** (a-d) LEED patterns from graphite/SiC(0001). The sample was heated several times to successively higher temperatures. (a) 1050 °C for 10 min. Immediately after oxide removal, showing SiC 1 × 1 pattern. (b) 1100 °C, 3 min. The √3 × √3 reconstruction is seen. (c) 1250 °C, 20 min. Pattern showing diffracted beams from the 6√3 × 6√3 unit cell. Examples of first-order SiC and graphite spots are marked. Note the surrounding hexagons of "6 × 6" spots. AES C:Si ratio 2:1 (~1 ML graphite). (d) 1400 °C, 8 min. AES ratio 7.5:1 indicates ~2.5 ML graphite. (e) STM image of a surface region. Inset: Atomically resolved region. (f): dI/dV spectra (log scale) acquired from the regions marked with corresponding line types in the image at top. The solid line is an average of 396 spectra at different positions, the dashed line an average of 105. Reproduced with permission from [78] Copyright 2004, ACS.

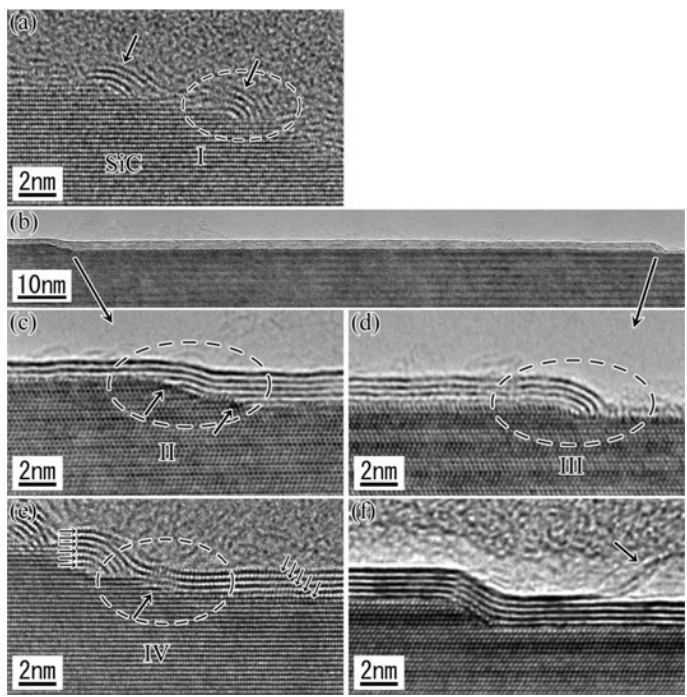

**Figure 13** HRTEM images of typical graphene layers formed on stepped SiC. (a) Initial growth occurs on steps. (b) Wide- area scan at a later stage of growth. The images in (c) and (d) are enlargements of part of (b). Images (e) and (f) show graphene around the step. Reproduced with permission from [85.a] Copyright 2010, ACS.

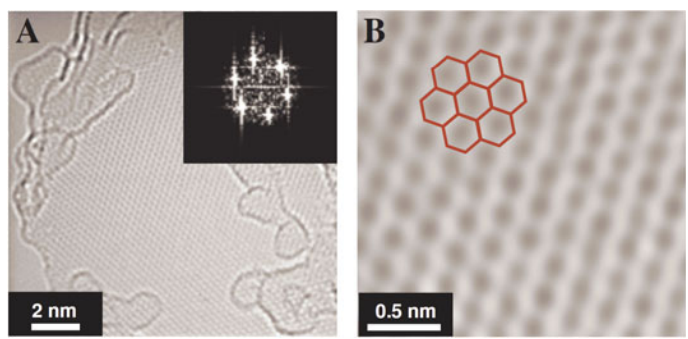

**Figure 14**. TEM images of graphene produced via the carbothermal reduction of SiOx to SiC.

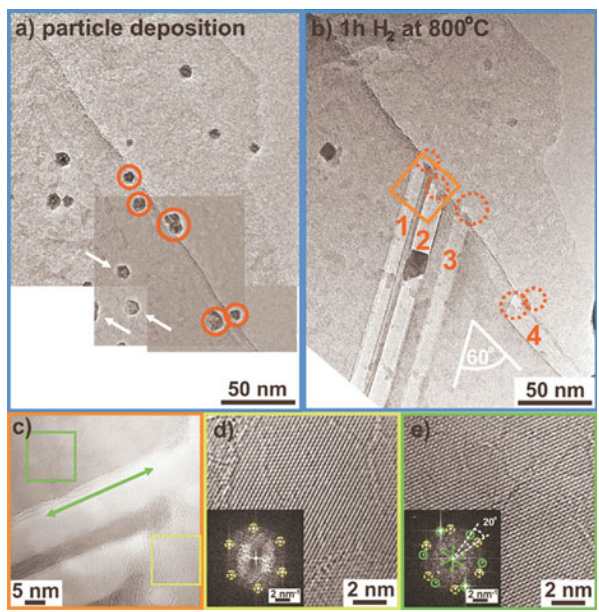

**Figure 15.** Use of robust perforated Si3N4 TEM grids enables one to monitor the different process steps of catalytic hydrogenation: (a) FLG flake after deposition of gas-phase-prepared Co nanoparticles; (b) FLG flake after catalytic hydrogenation at 800 °C for 1 h; (c) HRTEM micrograph of the area marked orange in panel b; the image is rotated clockwise by 50 ; (d,e) FFT-enhanced HRTEM micrographs and their FFTs (as insets) of the areas marked yellow and green in panel c, respectively. All circles in the FFTs resemble armchair directions in the graphite lattice. In the FFT of panel e, two sets of armchair directions are observed, which is indicative of a rotational stacking fault. The green arrows in panel e resemble the graphite zigzag directions with respect to the green circled armchair spots, indicating that the etching direction is a zigzag direction. Reproduced with permission from [120]. Copyright 2011, ACS.

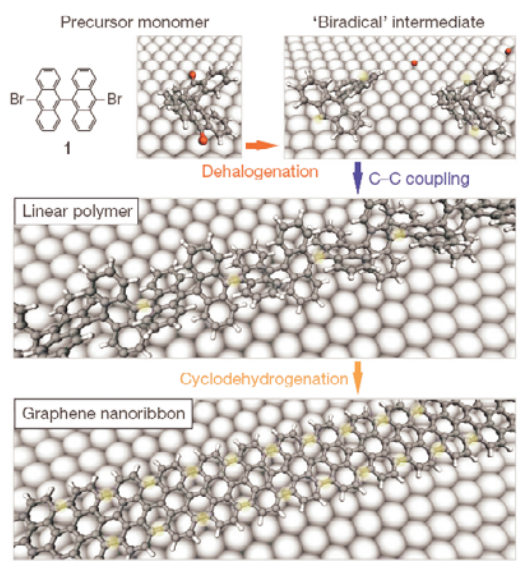

**Figure 16.** Bottom-up fabrication of atomically precise GNRs. Basic steps for surface-supported GNR synthesis, illustrated with a ball-and-stick model of the example of 10,109-dibromo-9,99-bianthryl monomers (1). Grey, carbon; white, hydrogen; red, halogens; underlying surface atoms shown by large spheres. Top, dehalogenation during adsorption of the di-halogen functionalized precursor monomers. Middle, formation of linear polymers by covalent interlinking of the dehalogenated intermediates. Bottom, formation of fully aromatic

GNRs by cyclodehydrogenation. Reproduced with permission from [122] Copyright 2010, NPG.

Authors Bios

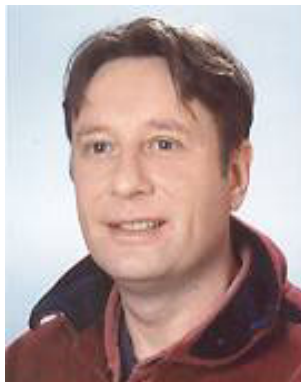

Mark H Rümmeli earned his PhD from London Metropolitan University. He then worked as a Research Fellow at the German Aerospace Center (DLR) in the Institute of Space Sensor Technology and Planetary Exploration. Currently he heads the Molecular Nanostructures group at the Leibniz Institute for Solid State and Materials Research Dresden. His research interests include understanding the growth mechanisms of nanostructures, in particular carbon based nanomaterials as well as advanced techniques for their functionalisation.

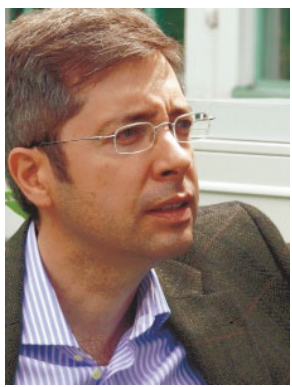

Professor Gianaurelio Cuniberti currently resides the Chair of Materials Science and Nanotechnology at the Dresden University of Technology and the Max Bergmann Center of Biomaterials Dresden. He studied Physics at the University of Genoa and at the University of Hamburg. He was a visiting scientist at MIT, the Max Planck Institute for the Physics of Complex Systems Dresden. From 2003 till 2007 he led a Volkswagen Foundation Junior Research Group at the University of Regensburg.

TOC entry

The future of graphene as a material for electronic based devices will depend heavily on our ability to piece graphene together as a single crystal and define its edges with atomic precision. In this progress report, current synthesis strategies for graphene and their weaknesses in terms of electronics applications are highlighted.

Keyword Graphene

Authors

Mark H. Rümmeli, Claudia G. Rocha, Frank Ortmann, Imad Ibrahim, Haldun Sevincli, Felix Börrnert, Jens Kunstmann, Alicja Bachmatiuk, Markus Pötsche, Masashi Shiraishi, Meyya Meyyappan, Bernd Büchner, Stephan Roche and Gianaurelio Cuniberti

Title

Graphene: Piecing it together

TOC figure

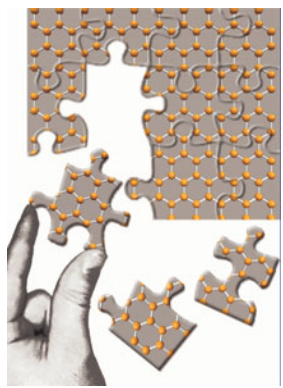